\def\year{2021}\relax
\documentclass[letterpaper]{article}
\usepackage{aaai21} %
\usepackage{times} %
\usepackage{helvet} %
\usepackage{courier} %
\usepackage[hyphens]{url} %
\usepackage{graphicx} %
\usepackage{graphicx} %
\usepackage{natbib}
\usepackage{caption}
\AtBeginDocument{%
  \providecommand\BibTeX{{%
    \normalfont B\kern-0.5em{\scshape i\kern-0.25em b}\kern-0.8em\TeX}}}

\usepackage{amsmath}

\usepackage{booktabs}

\usepackage{xcolor}
\definecolor{boxgrey}{HTML}{F3F3F3}

\usepackage{siunitx}

\newcommand{\hlbox}[2]{%
  \begin{center}%
    \fcolorbox{white}{boxgrey}{%
      \parbox{.9\columnwidth}{\noindent \textbf{#1}. \textit{#2}}
    }%
  \end{center}%
}

\usepackage{xspace}
\newcommand{\motiv}{\emph{MoTiV}\xspace}

\usepackage{newunicodechar}
\newunicodechar{»}{{\fontencoding{T1}\selectfont\guillemotright}}
\newunicodechar{«}{{\fontencoding{T1}\selectfont\guillemotleft}}

\usepackage{etoolbox}
\makeatletter
\patchcmd{\maketitle}{\copyright@text}{{\footnotesize %
ArXiV preprint\\
\copyright~\year~released under a Creative Commons license Attribution-ShareAlike 4.0 International (CC BY-SA 4.0)\\
\url{https://creativecommons.org/licenses/by-sa/4.0/}%
}}{}{}
\makeatother

\begin{document}

\title{What's Your Value of Travel Time? \\Collecting Traveler-Centered Mobility Data via Crowdsourcing}

\author{
Cristian Consonni,\textsuperscript{\rm 1}
Silvia Basile,\textsuperscript{\rm 1}
Matteo Manca,\textsuperscript{\rm 1}
Ludovico Boratto,\textsuperscript{\rm 1}
Andr\'e Freitas,\textsuperscript{\rm 2} \\
Tatiana Kovacikova,\textsuperscript{\rm 3}
Ghadir Pourhashem,\textsuperscript{\rm 3}
Yannick Cornet\textsuperscript{\rm 3}\\
}
\affiliations {
\textsuperscript{\rm 1}Data Science and Big Data Analytics, EURECAT - Centre Tecnol\`ogic de Catalunya - Barcelona (Spain),  \\
\textsuperscript{\rm 2}TIS.pt - Consultores em Transportes Inovação e Sistemas, S.A. - Lisbon (Portugal), \\
\textsuperscript{\rm 3}Dept. of International Research Projects - ERAdiate+, University of Žilina - Žilina (Slovakia) \\
cristian.consonni@acm.org, silvia.bas95@gmail.com, matteomanca@gmail.com, ludovico.boratto@acm.org,
andre.freitas@tis.pt,
\{tatiana.kovacikova,ghadir.pourhashem,yannick.cornet\}@uniza.sk\\
}

\maketitle
\begin{abstract}
Mobility and transport, by their nature, involve crowds and require the coordination of multiple stakeholders - such as policy-makers, planners, transport operators, and the travelers themselves. However, traditional approaches have been focused on time savings, proposing to users solutions that include the shortest or fastest paths. We argue that this approach towards travel time value is not centered on a traveler's perspective. To date, very few works have mined data from crowds of travelers to test the efficacy and efficiency of novel mobility paradigms. In this paper, we build upon a different paradigm of \emph{``worthwhile time''} in which travelers can use their travel time for other activities; we present a new dataset, which contains data about travelers and their journeys, collected from a dedicated mobile application. Each trip contains multi-faceted information: from the transport mode, through its evaluation, to the positive and negative experience factors. To showcase this new dataset's potential, we also present a use case, which compares corresponding trip legs with different transport modes, studying experience factors that negatively impact users using cycling and public transport as alternatives to cars. We conclude by discussing other application domains and research opportunities enabled by the dataset.
\end{abstract}

\maketitle

\section{Introduction}
The extraction of actionable knowledge from user mobility has been a central perspective mainly for mobility stakeholders - such as policy-makers, planners, and transport operators - who, in turn, used this knowledge to adapt the services to the end users. This process, known as behavioral-data mining, aims at extracting patterns from the travelers' behavior, in order to better characterize them~\cite{MancaBC18,BorattoCKM18}. Under this paradigm, end-users have mostly been passive actors. 

In the last 20 years, the integration between information and communication technologies (ICT) and transport have radically changed mobility: from the mere availability of information about transport such as bus lines or time tables, we now have new forms of offerings in terms of on-demand services and even whole new business models for transport, where new services offer the opportunity to share rides or even private cars. New Web platforms, such as journey planners~\cite{SourlasN19}, have transformed users into {\em active actors} in providing their mobility preferences. Users can sort trips based on different types of preferences (e.g., length, duration, emissions) and decide to complete a trip by combining services of different operators. Hence, novel mobility paradigms emerged, in which users' {\em value of travel time} (VTT) is defined as the combination of different factors (e.g., a trip by bike could be longer than taking other means of transport, but might valuable from multiple perspectives, such as emissions, costs, and fitness for the user)~\cite{DevarasettyBD12}.

However, other aspects of transportation, such as the reliance upon private cars have not changed. We argue that one of the main opportunities offered by the integration between ICT and transport lies in the possibility of gather data from a vast audience of travelers about both their travel preferences and wishes and their actual day-to-day transport usage.
In this sense, crowdsourcing, intended as the process of collecting data contributed by vast amounts of people, mainly via the Internet, can become a valuable means towards the accomplishment of this goal.

While novel paradigms related to value of travel time have been introduced~\cite{KaradimceLY18,KovacikovaLP18,LuganoKHP19}, existing studies capture user behavior in mobility from a single perspective. When moving from academic research to industrial applications, a one-to-one relationship between value of travel time and shortest path exists, so that existing services - such as Google Maps and Waze - mainly rank trips by shortest path. We believe that this might be due - at least from the research side - to the lack of knowledge on {\em what is actually valuable for the users in their mobility choices}. While several existing datasets capture mobility by considering a single perspective (e.g., one transport mode, or only trip coordinates), {\em no dataset capturing both explicit preferences in terms of value for the users when making their mobility choices and implicit information coming from the trip (e.g., coordinates) exist}. This, in turn, has been reflected in industrial applications, which do not capture explicit feedback on additional factors. Hence, value of travel time remains a concept trapped inside the mobility community, which cannot be fully exploited by the Web community, to create actionable knowledge for Web platforms. We believe that {\bf crowdsourcing can be a powerful tool to directly collect feedback on the travelers' value of travel time and their mobility choices.} %

To overcome these issues, in this paper we present the \motiv dataset, a collection of  traveler-centered information about trips and the value of travel time behind traveler's choices.

To show how the extraction of knowledge on users' value of travel time can concretely impact transport stakeholders, we also present a use case that characterizes what are the negative factors associated to a trip, when this is performed by means of public transport or by cycling, w.r.t.  the same trip performed by car.

This dataset was collected in the context of the Horizon 2020 \motiv (Mobility and Time Value) project, whose goal was to provide novel definitions of value of travel time (VTT)~\cite{MalichovaPKH20}.

\section{Related Work}\label{sec:related}

In this section, we present datasets related to the one presented in this paper, and conclude by highlighting the difference between our dataset and the existing ones. %

\subsection{User Behavior Mining} Data coming from online platforms was previously used to mine user behavior, to consider aspects such as their willingness to pay for services~\cite{ZografosAA12} and challenges in behavior change~\cite{SchrammelPMBT15}.  
\citeauthor{MancaBRGK17}~\cite{MancaBRGK17} presented a survey on mobility patterns considering social media data. \citeauthor{GonzalesHB08}~\cite{GonzalesHB08}, instead, consider mobile phone data, while \citeauthor{CalabreseDDFR13}~\cite{CalabreseDDFR13} extract mobility patterns from urban sensing data. \citeauthor{Goulias18}~\cite{Goulias18} surveyed the existing travel behavior models.
Data analysis can produce insights that serve as input for other purposes, such as the improvement of transport services by considering user needs~\cite{SierpinskiS17}, the promotion of changes of the user habits~\cite{SchrammelPMBT15}, and the improvement of journey planners and transport portals~\cite{Esztergar16,VargasWR11}. 
Other studies go beyond data analysis, e.g., to extract topic models from geo-location data~\cite{HasanU14}, to forecast the evolution of preferences over time thanks to a Markov model~\cite{ZarwiVW17}, or to provide a personalized journey planning~\cite{JakobHORZF14}.
As previously mentioned, the knowledge coming from the user analytics can also be used as a form of actionable knowledge, e.g., to improve transport service according to the user needs~\cite{SierpinskiS17}, to promote changes in the user habits~\cite{SchrammelPMBT15} such as the adoption of greener and healthier solutions~\cite{GabrielliFJWSHNMMJ14}, or improve the usability and services provided by journey planners and transport portals~\cite{Esztergar16,VargasWR11}.

\subsection{Crowdsourced Datasets}

\paragraph{Social and trip data} Microsoft GeoLife is a social network that also allows users to share their experience, both through GPS data and with pictures. The publicly available GPS trajectory dataset was collected by 182 users, for over two years\footnote{\url{https://www.microsoft.com/en-us/research/project/geolife-building-social-networks-using-human-location-history/#!downloads}}. No explicit info about the means of transport is available, and no evaluation of the trips is offered.

\paragraph{Check-in data} Check-in data, coming from platforms such as Twitter, Foursquare, or Gowalla, is usually used to consider user preferences related to their mobility. Indeed, knowing where the users go and with which frequency, allows to characterize users' mobility and their preferences (especially if check-ins can be paired with reviews). Examples of datasets belonging to this class can be found here~\footnote{\url{https://sites.google.com/site/yangdingqi/home/foursquare-dataset}}. While this class of datasets is widely employed in various personalization algorithms, the concept of a trip is entirely missing, thus losing all the information about the means of transport, or the relevant factors for the users.

\paragraph{Trip-only data} Another class of datasets collect trip information. A selection of 250 datasets available for research purposes can be found here\footnote{\url{https://data.world/datasets/trips}}. Trip datasets usually do not associate the trips to a user, thus not allowing a characterization of the individual user mobility. In addition, they are usually associated to a single transport of mode (e.g., taxis).

\subsection{Contextualizing our Contributions} As this analysis has presented, mobility datasets usually {\em capture a single perspective} and usually the full user-mobility picture as lost. The aim of our data collection is to capture multiple perspectives behind user mobility. This is done by: 
\begin{itemize}
    \item having trips with different transport modes (something not available in classic trip datasets);
    \item collecting trip coordinates (this is usually not available in check-in datasets);
    \item collecting reviews and evaluations of each trips;
    \item working at a lower granularity, by not only considering the trip as a whole, but also the activities that compose it; This allows to consider the different experience factors in user mobility, at different stages of the trip.
\end{itemize}

\section{Background and Crowdsourcing Process}

\hlbox{Worthwhile time}{Worthwhile time is a central concept for our data collection. While the traditional view is to consider travel time as something to minimize, we consider travel time as an opportunity, i. e. time that that can be characterized by other activities. We have collected data through a dedicated mobile app, called \emph{Woorti}\footnotemark, to ground the definition of worthwhile time into multiple dimensions.}\footnotetext{The name of the app is a play on the words ``worth it'' referring to worthwhile travel time.}

A key challenge of collecting a European-wide dataset was engaging users across multiple countries for an extended time. In fact, from a traveler's point-of-view, the data collection translated to using the smartphone app actively for at least 14 days. We framed the process as a challenge between user: data-collection campaigns (DCCs) would group users together and reward the most active ones in each campaign. To facilitate the process, we have partnered with thematic organizations across Europe - such as cycling associations - and we have appointed DCC managers as liaison with the community, whose tasks were recruiting participants, promoting and monitoring the data collection process. The application supported both Android and iOS devices and was available in 11 languages. Data collection targeted 10 European countries: Belgium, Croatia, Finland, France, Italy, Norway, Portugal, Slovakia, Spain, and Switzerland. Overall, the data considered in this paper cover a period of 8 months, from May 1st, 2019 to December, 13rd 2019.

The use of the app consists of three main phases:
\begin{enumerate}
    \item \emph{Onboarding}: upon installing the application and registering a new account, the user is introduced to the functionalities of the app. During this process, the user enters their travel preferences as well as some basic demographic information.
    \item \emph{Trip recording}: the user can start a new trip and the app automatically collects data in background.
    \item \emph{Trip validation}: when a trip is finished, the user can review the data, validate it and insert other data regarding the trip (trip purpose, mood, etc). When validating a trip, the user must choose one leg of the trip as the reviewed leg.
\end{enumerate}
Within the application, a user can access their data and visualize and edit their profile information and trips.  Furthermore, the application features a dashboard that presents to the user statistics related to their validated trips.

The value of travel time is analyzed from a traveler’s perspective, assuming that time and cost savings are not always the main criteria influencing route and mode choice. Depending on the traveler's transport attitude and context, other criteria such as environmental impact, comfort, or even weather conditions may influence the perceived value of a trip. In particular, we adopt the perspective that travel time can be worthwhile, i. e. it can be allocated for activities that the user finds useful, enjoyable, or productive. For this reason, when validating a trip, the user is asked about which activities they have performed during the trip, which factors in their have influenced the trip positively or negatively, and which was the trip purpose.

By collecting this data, we are able to shifts perspective from considering travel time as spent - or, worse, wasted - to time that can be characterized by other activities. Furthermore, this characterization is not limited by defining time as productive or unproductive time, because it is not necessarily related to its evaluation in terms of cost. Worthwhile time is independent of what can be monetized.

The \motiv dataset and the analyses presented here contain only validated trips. During data collections, local campaign managers stressed to users that validating a trip meant sharing it with the project, i. e. validated trips were uploaded to a central server. This process preserves users' privacy since they were always in control of what data they are willing to share with the project. Furthermore, the data collection process was overseen and approved by an independently appointed ethics advisor and the ethics committee at the University of \u{Z}ilina, \motiv's coordinator.

\section{Dataset Description}\label{sec:dataset}

\hlbox{Dataset}{%
The dataset can be downloaded at: \\ \url{https://zenodo.org/record/4027465}. \\
The code used for pre-processing the raw data and performing the case study is available at: \\ \url{https://github.com/MoTiV-project/data-analysis}
}

\noindent{}The dataset is comprised of $13$ tables saved as comma-separated value (\texttt{.csv}) files:%
\begin{enumerate}
  \item \path{user_details.csv}
  \item \path{user_generic_worthwhileness_values.csv}
  \item \path{user_specific_worthwhileness_values.csv}
  \item \path{mots.csv}
  \item \path{trips.csv}
  \item \path{legs.csv}
  \item \path{legs_coordinates.csv}
  \item \path{activities.csv}
  \item \path{experience_factors.csv}
  \item \path{purposes.csv}
  \item \path{weather_legs.csv}
  \item \path{weather_raw.csv}
  \item \path{worthwhileness_elements_from_trips.csv}
\end{enumerate}%

\paragraph{User Details}
The table \path{user_details.csv} $(1)$ contains the data collected during the onboarding phase. Only the fields \texttt{user\_id}, \texttt{registration\_timestamp}, \texttt{gender}, and \texttt{age} are required.

\begin{itemize}
  \item \texttt{user\_id}: the user identifier, a string of 29 characters.
  \item \texttt{registration\_timestamp}: the trip start date, formatted as \texttt{\%Y-\%m-\%d \%H:\%M:\%S.\%f}.
  \item \texttt{gender}: the gender of the user. Three values are possible: \texttt{Other}, \texttt{Male}, \texttt{Female}.
  \item \texttt{age}: the user's age in a range. Possible values are: \texttt{16-19}, \texttt{20-24}, \texttt{25-29}, \texttt{30-39}, \texttt{40-49}, \texttt{50-64}, \texttt{65-74}, \texttt{75+}.
  \item \texttt{language}: the language used within the app expressed as a 3-letter ISO 639-2/B code.
  \item \texttt{city}: the user's city of residence.
  \item \texttt{country}: the user's country of residence expressed in ISO 3166-1 alpha-3 code.
  \item \texttt{education\_level}: the user's education, possible values are: \texttt{Basic (up to 10th grade)}, \texttt{High School (12th grade)}, and \texttt{University}.
  \item \texttt{marital\_status}: the user's marital status, possible values are: \texttt{Single}, \texttt{Registered relationship}, \texttt{Married}, \texttt{Divorced}, \texttt{Widowed}.
  \item \texttt{number\_of\_people}: the number of people in the household, it can be a number from $1$ to $4$ or $5+$.
  \item \texttt{labour\_status}: the users' labour status. Possible values are: \texttt{Student}, \texttt{Employed full-time}, \texttt{Employed part-time}, \texttt{Pensioner}, \texttt{Unemployed}.
  \item \texttt{years\_of\_residence}: the number of years of residence in the household. Possible values are: \texttt{Less than 1}, \texttt{1 to 5}, \texttt{More than 5}.
\end{itemize}

\paragraph{User's Generic and Specific Worthwhileness Elements}
Users' preferences and experiences are characterized along the three dimensions of fitness, enjoyment, and productivity, defined as follows (in parenthesis, we report a description of each dimension, visualized by users during data collection):

\begin{itemize}
    \item fitness measures how much the user values the fact that when traveling they can exercise («When you walk, cycle, or even run on your travels, you are getting exercise and keeping in shape»);
    \item enjoyment is related to how the travel can be used for fun or relaxing activities («Relaxing or having fun: taking time to listen to music, rest or meditate; engaging in social media; observing the surroundings»);
    \item productivity captures how much the user values the possibility of using travel time to complete some tasks, either personal or work-related. («Using travel time to get things done, not only for work or study, but also personal things like managing home or family stuff»). It is further diving in two aspects: \emph{Paid work} and \emph{Personal tasks}.
\end{itemize}

User preferences and experiences are encoded in two main sets of values, called worthwhileness\footnote{Although this diction of the word is less widespread than the more common variant ``\emph{worthiness},'' it is used throughout the project, so we keep it for consistency with the project itself.} values:
\begin{itemize}
    \item \emph{generic worthwhileness values:} they are a triplet of values $(F, E, P)$ for \emph{fitness}, \emph{enjoyment}, and \emph{productivity}, respectively. They measure how much the user values these dimension in general when traveling;
    \item \emph{specific worthwhileness values:} they are triplets of values $(F, E, P)$ that the user is asked to assign for each specific mode of transport chosen in the onboarding phase. The transport modes that the user selects during the onboarding phase are called \emph{preferred transport modes}. Specific worthwhileness values are the measure of how much the user values \emph{fitness}, \emph{enjoyment}, and \emph{productivity} when using that particular transport mode.
\end{itemize}
During the onboarding phase, the user is asked to provide both the generic and specific worthwhileness values on a scale from 1 to 100. When evaluating trips, the user is asked to provide an evaluation for each dimension of \emph{fitness}, \emph{enjoyment}, and \emph{productivity} using a scale from low to high (\texttt{low}, \texttt{medium}, \texttt{high}). This difference in data collection depended on the design of the app, whose description is out of scope for this paper. For consistency with the evaluation values, we scale the onboarding values to the same three classes: \texttt{low}, for values in $[0-33]$; \texttt{medium}, $[34-66]$; and \texttt{high}, $[67-100]$.

Generic worthwhileness values are stored in the table \path{user_generic_worthwhileness_values.csv} $(2)$, its columns are:
\begin{itemize}
    \item \texttt{userid}, the user identifier;
    \item \texttt{fit}, the value for fitness $[0-100]$;
    \item \texttt{prod}, the value for productivity $[0-100]$;
    \item \texttt{enjoy}, the value for enjoyment $[0-100]$;
\end{itemize}

Specific worthwhileness elements are related to specific mode of transports, chosen by the user during the \emph{Onboarding} phase. These are referred as the traveler's favorite modes of transport. Specific worthwhileness values are stored in the table \path{user_specific_worthwhileness_values.csv} $(3)$. Its columns are:
\begin{itemize}
    \item \texttt{userid}, the user identifier;
    \item \texttt{motid}, the mode of transport identifier;
    \item \texttt{fit}, the value for fitness $[0-100]$;
    \item \texttt{prod}, the value for productivity $[0-100]$;
    \item \texttt{enjoy}, the value for enjoyment $[0-100]$;
\end{itemize}
The mapping between mode of transport ids and text is contained in \path{mots.csv} $(4)$.

\begin{table*}[h!]
\centering
\small
\clearpage{}%
\begin{tabular}{@{}p{4.5cm}p{12.5cm}@{}}
\toprule
\textbf{Field}
                                  & \textbf{Description and admissible values} \\ \hline
\texttt{tripid}
                                  & Trip identifier, a string in the format \texttt{\#\{n1\}:\{n2\}} where \texttt{n1} and \texttt{n2} are two numbers. \\
\texttt{userid}
                                  & User identifier. \\
\texttt{start\_date}
                                  & Trip start date, formatted as '\%Y-\%m-\%d \%H:\%M:\%S.\%f'. \\
\texttt{end\_date}
                                  & Trip end date, formatted as '\%Y-\%m-\%d \%H:\%M:\%S.\%f'. \\
\texttt{average\_speed}
                                  & Average speed during trip in km/h. \\
\texttt{max\_speed}
                                  & Max speed during trip. \\
\texttt{distance}
                                  & Leg distance in meters. \\
\texttt{duration}
                                  & Leg duration in seconds. \\
\texttt{mood\_rating}
                                  & Evaluation of trip mood on a scale from 1 to 5. \\
\texttt{did\_you\_have\_to\_arrive}
                                  & Answer to the question \emph{``Did you have to arrive on time''}, a boolean value. \\
\texttt{how\_often}
                                  & Answer to the question \emph{``How often do you make this trip?''}, possible values are integers from \texttt{0} to \texttt{3}. \texttt{-1} means that the question was not answered. \\
\texttt{use\_trip\_more\_for}
                                  & Answer to the question \emph{``How often do you make this trip?''}, possible values are integers from \texttt{0} to \texttt{3}. \texttt{-1} means that the question was not answered. \\
\texttt{manual\_start}
                                  & Answer to the question \emph{``Has the trip recording been started manually?''}, a boolean value. \\
\texttt{manual\_end}
                                  & Answer to the question \emph{``Has the trip recording been ended manually?''}, a boolean value. \\
\texttt{validation\_date}
                                  & Trip validation date, formatted as '\%Y-\%m-\%d \%H:\%M:\%S.\%f'. \\
\texttt{os}
                                  & Operating system of the phone, can be \texttt{iOS} or \texttt{Android}, a string.  \\
\texttt{os\_version}
                                  & Operating system version, a string. \\
\texttt{model}
                                  & Model of the phone, a string. \\  
\bottomrule
\end{tabular}\clearpage{}%

\caption{Trip information stored in \texttt{trips.csv}.}
\label{tab:trips}
\end{table*}

\begin{table*}[h!]
\centering
\small
\clearpage{}%
\begin{tabular}{@{}p{4.5cm}p{12.5cm}@{}}
\toprule
\textbf{Field}
          & \textbf{Description and admissible values} \\ \midrule
\texttt{legid}
          & Leg identifier, a string in the format \texttt{\#\{n1\}:\{n2\}} where \texttt{n1} and \texttt{n2} are two numbers. \\
\texttt{class}
          & Leg type, either \texttt{Leg} or \texttt{WaitingEvent}. \\
\texttt{userid}
          & User identifier. \\
\texttt{tripid}
          & Trip identifier. \\
\texttt{motid}
          & Mode of transport identifier, an integer number. \\
\texttt{start\_date}
          & Leg start date, formatted as \texttt{\%Y-\%m-\%d \%H:\%M:\%S.\%f}. \\
\texttt{end\_date}
          & Leg end date, formatted as \texttt{\%Y-\%m-\%d \%H:\%M:\%S.\%f}. \\
\texttt{true\_distance}
          & Leg distance in meters. \\
\texttt{leg\_distance}
          & Leg distance in meters. \\
\texttt{leg\_duration}
          & Leg duration in seconds. \\
\texttt{worthwhileness\_rating}
          & Worthwhileness rating, $-1$ if not set. \\
\texttt{transport\_category}
          & Transport category. \\
\texttt{campaign}
          & Data collection campaign. \\
\texttt{weekday}
          & Day of the week when the leg was performed. \\
\texttt{weekday\_class}
          & Classification of the day, either \texttt{Working\_day} or \texttt{Weekend}. \\
\bottomrule
\end{tabular}\clearpage{}%

\caption{Leg information stored in \texttt{legs.csv}.}
\label{tab:legs}
\end{table*}

\begin{table*}[ht!]
\centering
\small
\clearpage{}%
\begin{tabular}{@{}p{3.0cm}p{14.0cm}@{}}
\toprule
\textbf{Field}          & \textbf{Description and admissible values}                                                       \\ \hline                                                       
\texttt{leg\_id}                              & The identifier of the leg in the format \#n1:n2, with n1 and n2 being two integer numbers.                                                                                                             \\ 
\texttt{start\_lat}                           & a real number with 3 decimals. The latitude of the starting point of the leg.                                                                                                                          \\ 
\texttt{start\_lon}                           & a real number with 3 decimals. The longitude of the starting point of the leg.                                                                                                                         \\ 
\texttt{end\_lat}                             & a real number with 3 decimals. The latitude of the ending point of the leg.                                                                                                                            \\ 
\texttt{end\_lon}                             & a real number with 3 decimals. The longitude of the ending point of the leg.                                                                                                                           \\ 
\texttt{start\_name}                          & The inferred name of the city where the starting point of the leg is located. This field is obtained in the preprocessing by integrating the leg coordinates with OECD data.                           \\ 
\texttt{start\_country}                       & The ISO 3166 Alpha-3 code of the inferred country where the starting point of the leg is located. This field is obtained in the preprocessing by integrating the leg coordinates with OECD data1b.     \\ 
\texttt{start\_class}                         & Possible values are: urban, sub-urban, or rural. The classification of the starting point of the leg. This field is obtained in the preprocessing by integrating the leg coordinates with OECD data1c. \\ 
\texttt{end\_name}                            & The inferred name of the city where the ending point of the leg is located. This field is obtained in the preprocessing by integrating the leg coordinates with OECD data.                             \\ 
\texttt{end\_country}                         & The ISO 3166 Alpha-3 code of the inferred country where the ending point of the leg is located. This field is obtained in the preprocessing by integrating the leg coordinates with OECD data1b.       \\ 
\texttt{end\_class}                           & Possible values are: urban, sub-urban, or rural. The classification of the ending point of the leg. This field is obtained in the preprocessing by integrating the leg coordinates with OECD data1c.   \\ \bottomrule
\end{tabular}\clearpage{}%

\caption{Legs coordinates stored in  \texttt{legs\_coordinates.csv}}
\label{tab:legs_coords}
\end{table*}

\paragraph{Trips Info}
Trip data are contained in the table \path{trips.csv} $(5)$, whose fields are reported in Table~\ref{tab:trips}. A trip is a collection of \emph{legs} and \emph{waiting events}: the former are parts of a journey where the app has detected some movement\footnotetext{The app offers mode detection algorithms to infer the type of mode of transport used; however during the trip validation ad evaluation the user was able to override the modes suggested.}, while waiting events are intervals of time where the app did not detect any significant displacement. We processed trip data to merge legs that were erroneously split by the app, furthermore we performed outlier detection to eliminate the trips that were in top and bottom percentiles in terms of length and duration. 

\paragraph{Legs Info}
Leg data are contained in the table \path{legs.csv} $(6)$, whose fields are reported in Table~\ref{tab:legs}. 

\paragraph{Legs Coordinates}
Leg coordinates are contained in the table \path{legs_coordinates.csv} $(7)$, whose fields are reported in Table~\ref{tab:legs_coords}. Coordinates are anonymized depending on the fact that the point is located in a urban, suburban or rural area. To classify the points we used the ``functional urban areas by country'' classification provided by the Organisation for Economic Co-operation and Development (OECD).\footnote{Functional urban areas by country, \url{https://www.oecd.org/regional/regional-statistics/functional-urban-areas.htm}}

Points were anonymised using the following criterion, applied both to latitude and longitude: 
\begin{itemize}
    \item for urban areas, points are rounded to the third decimal place; 
    \item for sub-urban areas, the third decimal place is rounded to the nearest $0.5$;
    \item for rural areas, the second decimal place is rounded to the nearest $0.5$;
\end{itemize}

\paragraph{Activities}
Data about activities performed during trips are contained in table \path{activities.csv} $(8)$. A user can select multiple activities during a trip, but we do not record their duration. Columns in \path{activities.csv} are: 

\begin{itemize}
    \item \texttt{legid}, leg identifier
    \item \texttt{activity}, activity name, possible values are \texttt{Accompanying}, \texttt{Browsing}, \texttt{Cycling}, \texttt{Driving}, \texttt{Eating}, \texttt{Listening},\\ \texttt{PersonalCare}, \texttt{ReadingDevice}, \texttt{ReadingPaper}, \texttt{Relaxing}, \texttt{Talking}, \texttt{Thinking}, \texttt{Walking}, \texttt{Watching}, and \texttt{Other}
\end{itemize}

\paragraph{Experience Factors}
Table \path{experience_factors.csv} $(9)$ contains information about experience factors affecting trips, its columns are:

\begin{itemize}
    \item \texttt{legid}, leg identifier;
    \item \texttt{factor}, experience factor name. There are 48 possible values, for context we list here a sample:
          \texttt{Air\_Quality}, \texttt{Cleanliness},  \texttt{Crowdedness\_Seating}, \texttt{Internet\_Connectivity}, 
          \texttt{Privacy}, \texttt{Reliability\_Of\_Travel\_Time}, \texttt{Todays\_Weather}, \texttt{Toilets}, 
          \texttt{Vehicle\_Quality}, \texttt{Vehicle\_Ride\_Smoothness};
    \item \texttt{type}, factor categorization;
    \item \texttt{minus}, a boolean value, the factor was rated as negative;
    \item \texttt{plus}, a boolean value, the factor was rated as positive.
\end{itemize}

\paragraph{Trip purposes}
The table \path{purposes.csv} $(10)$ contains information about trips purposes, its columns are: 

\begin{itemize}
    \item \texttt{tripid}, the trip identifier;
    \item \texttt{purpose}, the trip purpose, possible values are \\ \texttt{Business\_Trip}, \texttt{Everyday\_Shopping}, \texttt{Home}, \\ \texttt{Leisure\_Hobby}, \path{Personal_Tasks_Errands}, \\ \texttt{Pick\_Up\_Drop\_Off}, \texttt{Work}, and \texttt{Other}.
\end{itemize}

\paragraph{Weather data}
Weather data has been collected through the API provided by OpenWeatherMap.\footnote{\url{https://openweathermap.org/}, OpenWeatherMap is an online service owned by OpenWeather Ltd.} The API was queried regularly for a set of $66$ cities of interest in the scope of the project. Weather information was collected for the times of 09:00, 12:00 and 18:00 for each day from July 8th, 2019 to December 18th, 2019. The dataset contains two tables related to weather data: \path{weather_raw.csv} $(11)$ and \path{weather_legs.csv} $(12)$.

The table \path{weather_raw.csv} contains the data parsed has obtained from the OpenWeather API, the documentation of each field is available on the OpenWeatherMap website.\footnote{OpenWeatherMap historical weather API guide, \url{https://openweathermap.org/history}.} The table \path{weather_legs.csv} contains the association between trips legs and the corresponding weather for the time and place, the available fields are presented in Table~\ref{tab:weather_legs}.

\begin{table*}[ht!]
\centering
\small
\clearpage{}%
\begin{tabular}{@{}p{5.0cm}p{12.0cm}@{}}
\toprule
\textbf{Field}
          & \textbf{Description and admissible values} \\ \midrule
\texttt{weatherid}
          & Weather identifier, a string of 24 characters. \\
\texttt{legid}
          & Leg identifier. \\
\texttt{request\_date}
          & Weather request timestamp, formatted as '\%Y-\%m-\%d \%H:\%M:\%S.\%f'. \\
\texttt{centroid\_x}
          & Longitude of the centroid of the leg. \\
\texttt{centroid\_y}
          & Latitude of the centroid of the leg. \\
\texttt{country}
          & Country the leg is located in. \\
\texttt{weather\_scenario}
          & Classification of weather scenarios. \\
\texttt{apparent\_temperature}
          & Apparent temperature, computed using Steadman's equation. \\
\texttt{net\_radiation}
          & Net radiation received by the terrain in the location of the centroid. \\
\texttt{temperature\_category}
          & Categorization based on apparent temperature. \\
\texttt{temperature\_description}
          & Description of the temperature based on apparent temperature. \\
\texttt{cloud\_category}
          & Categorization based on cloud cover. \\
\texttt{cloud\_main}
          & Description based on cloud cover. \\
\texttt{precipitation\_category}
          & Categorization based on precipitation level. \\
\texttt{precipitation\_main}
          & Description based on precipitation level. \\
\texttt{wind\_beaufort\_number}
          & Categorization using Beaufort's number, based on wind speed. \\
\texttt{wind\_category}
          & Categorization based on wind speed. \\
\texttt{wind\_description}
          & Description based on wind speed. \\
\bottomrule
\end{tabular}
\clearpage{}%

\caption{\label{tab:weather_legs}Weather info stored in \texttt{weather\_legs.csv}.}
\end{table*}

For the computation of weather scenarios, different sources of data were combined. %
Specifically, we computed the apparent temperature, which is equivalent of the temperature perceived by humans, based on effects of air temperature, relative humidity and wind speed. The formula for apparent temperature introduced by Robert Steadman in 1984~\cite{steadman1984universal} was used, which takes into consideration four
environmental factors: wind, temperature, humidity and radiation from the sun. %

\paragraph{Worthwhileness elements from trips}
Table \path{worthwhileness_elements_from_trips.csv} $(13)$ contains the evaluation of each trip along the worthwhileness elements dimensions on scale from \texttt{0} (low) to \texttt{2} (high). The fields of the table are:
\begin{itemize}
    \item \texttt{tripid}, the trip identifier;
    \item \texttt{legid}. the leg identifier;
    \item \texttt{worthwhileness\_element}, name of the wortwhileness element. Possible values are \texttt{Enjoyment}, \texttt{Fitness}, \texttt{Paid\_work}, \texttt{Personal\_tasks} and \texttt{Unknown};
    \item \texttt{value}, the value of the wortwhileness element, possible values are: \texttt{0}, \texttt{1}, \texttt{2} and \texttt{-1};
\end{itemize}
Records where the wortwhileness element and the value is \texttt{-1} are the same.

\subsection{A User's Story}
In the following section, we present a brief user story of a user using the dedicated app to describe the data collected in context. We will call our user Luigi. 

One day, Luigi sees an advertisement of a new app just released: \emph{Woorti}. \emph{Woorti} keeps track of personal trips along with all the aspects that influenced the travel, it seems interesting! So he decides to download it. He registers to the app and inserts some demographic information (\textbf{\path{user_details}}). After that, he is asked how much he values three aspects of life while he travels: enjoyment, how much fun he has; productivity, if he is able to work or do personal tasks; and fitness, if he is able to exercise and stay healthy; both in general terms (\textbf{\path{user_generic_worthwhileness_values}}) and with respect to his favorite transport modes (\textbf{\path{user_specific_worthwhileness_values}}).

Luigi wants to try the app while going to work (\textbf{\path{purposes}}). He starts from his apartment, walks until the bus stop where he waits for 5 minutes the bus. He gets on the bus for 20 minutes and in the meantime he reads a book and use the smartphone to check something on internet (\textbf{\path{activities}}). He easily finds a seat and enjoys his time on the bus since it is not crowded and he likes to watch the landscape outside (\textbf{\path{experience_factors}}). The bus stop is just few minutes from his office and since it's a beautiful and sunny day (\textbf{\path{weather}}) he decides to walk until there. Before starting his working day, he stops at the bar to take a coffee and spends some minutes filling the questionnaire on the app after his trip. He validates the trip checking that the start and end points were correct (\textbf{\path{legs_coordinates}}): he did in total four legs during his trip (walking, waiting event, bus, walking) (\textbf{\path{legs}}). He decides to review his leg on the bus and he completes all the questions, giving also a rate for the whole trip and for the leg he reviewed (\textbf{\path{trips}}), included a score for the enjoyment, productivity and fitness (\textbf{\path{worthwhileness_elements_from_trips}}). 

\section{Case Study}\label{sec:usecase}

\hlbox{Experience Factors Impacting Negatively Use of Public Transport, Cycling and Walking versus Private Cars}{We leverage the experience factor evaluations to compare trips done by cars versus trips conducted with other modes of transport: public transport, cycling and other emerging micromobility, and walking. With this use case we want to answer the question: ``What are the negative experience factors of cyclists and users of public transport for the same trip legs performed by car?''.}

We find users that have traveled a given route multiple times using both cars and alternative modes of transport and look at which are the factors that have impacted negatively the travel experience when using bikes or public transport. In this way, we want to get some  insights on which are the experience factors that are hindering the use of modes of transports alternative to cars. This analysis is performed along the following steps: 
\begin{enumerate}
    \item we select trip legs performed by car and those users that have performed at least one such trips;
    \item we restrict this set to the users that have chosen at least one preferred transport mode within the transport categories: \emph{cycling and emerging micromobility}, \emph{public transport (short distance)}, and \emph{public transport (long distance)}. For the sake of the study we call these the alternative transport category
    \item we identify users that have performed similar trips using different transport modes; specifically, we select users that performed the same journey using a car - in the \emph{private motorized} category - or using a bike or public transport.
    \item For the same user, we look at negative experience factors for trip legs performed using modes of transports in the alternative categories.
\end{enumerate}

\begin{figure}[h]
  \centering
  \includegraphics[width=0.75\columnwidth]{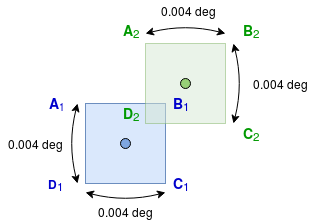}
  \caption{\label{fig:similar_points}Procedure to identify close points in space: each point is transformed to a curved square, where each side is an arc - with sides of length $0.004$ degrees. If the squares of two points intersect, we consider them a matching pair.}
\end{figure}

For the definition of similar trips we match trips using their starting and ending points: When using different modes of transport can lead to taking different paths, but for our analysis we are interested in the fact that a user needs to travel from a given pair of locations. To identify points that are close in space we adopt the following procedure, illustrated in Figure \ref{fig:similar_points}: each point is transformed to a curved square, where each side is an arc of length $0.004$ degrees. If the squares representing two points intersect, we consider them a matching pair. To simplify the computational complexity of the matching process we only compare trip legs in the same country. To estimate the distance between these two point we use an approximate conversion between the precision of decimal degrees in the EPSG:7030/WGS 84,\footnote{\url{http://epsg.io/7030-ellipsoid}, taken as the E/W at 45 degrees N/S.} $0.001$ degrees correspond $78.71 \,\mathrm{m}$\, so at most two points are matching if they are within a maximum distance of $445 \,\mathrm{m}$:
\[
    D = \sqrt{2} \cdot 0.004 \,\mathrm{deg} \cdot 78.71 \,\mathrm{km/deg} \approx 445 \,\mathrm{m}
\]

\begin{table}[b]
\centering
\small
\clearpage{}%
\begin{tabular}{@{}lp{2.8cm}rrr@{}}
\toprule
\textbf{Step} & \textbf{Step description} & \textbf{Trips} & \textbf{Legs} & \textbf{Users} \\ \midrule
1          & All trips, legs, users
                                          &       $64,098$ &     $158,897$ &        $3,269$ \\
2          & Users that have performed at least one trip leg by car
                                          &       $21,764$ &      $62,227$ &        $2,083$ \\
3          & Users that have selected at least one preferred alternative transport mode
                                          &       $51,973$ &     $132,259$ &        $1,771$ \\
4          & Users that have performed at least one trip leg by car and by alternative transport modes
                                          &       $20,032$ &      $27,921$ &        $1,376$ \\ \bottomrule
\end{tabular}\clearpage{}%

\caption{User statistics  considered by our case study.}
\label{tab:case-study}
\end{table}

\begin{table*}[h!]
\centering
\small
\clearpage{}%
\begin{tabular}%
  {@{}%
   >{\raggedright\arraybackslash}p{4cm}%
   >{\raggedleft\arraybackslash}p{1cm}%
   >{\raggedright\arraybackslash}p{4cm}%
   >{\raggedleft\arraybackslash}p{1cm}%
   >{\raggedright\arraybackslash}p{4cm}%
   >{\raggedleft\arraybackslash}p{1cm}%
   @{}%
  }
\toprule
  \multicolumn{2}{c}{\textbf{Cycling and micromobility}}
& \multicolumn{2}{c}{\textbf{Public transport (short distance)}}
& \multicolumn{2}{c}{\textbf{Public transport (long distance)}} \\
  \textbf{Factor}                  & \textbf{\#} 
& \textbf{Factor}                  & \textbf{\#}
& \textbf{Factor}                  & \textbf{\#} \\
\cmidrule{1-2}\cmidrule{3-4}\cmidrule{5-6}
Cars other vehicles                & $1,751$ & Privacy                        & $500$ & Internet connectivity          & $53$ \\
Air Quality                        & $1,277$ & Crowdedness seating            & $499$ & Privacy                        & $49$ \\
Road path availability and safety  & $1,219$ & Other people                   & $463$ & Seating quality personal space & $44$ \\
Noise level                        & $1,097$ & Seating quality personal space & $445$ & Reliability of travel time     & $43$ \\
Road path quality                  &   $969$ & Noise level                    & $437$ & Noise level                    & $36$ \\
Traffic signals crossings          &   $854$ & Internet connectivity          & $418$ & Other people                   & $35$ \\
Crowding congestion                &   $837$ & Charging opportunity           & $406$ & Vehicle ride smoothness        & $34$ \\
Today's weather                    &   $591$ & Air quality                    & $349$ & Today's weather                & $34$ \\
Simplicity difficulty of the route &   $417$ & Scenery                        & $309$ & Crowdedness seating            & $33$ \\
Facilities shower lockers          &   $371$ & Reliability of travel time     & $290$ & Food drink available           & $31$
\\ \bottomrule
\end{tabular}\clearpage{}%

\caption{Overall top-ten negative experience factors for all trip legs in the categories cycling and emerging micromobility, public transport (short distance), and public transport (short distance).}
\label{tab:all_experience_factors}
\end{table*}

\begin{table*}[h!]
\centering
\small
\clearpage{}%
\begin{tabular}%
  {@{}%
   >{\raggedright\arraybackslash}p{4cm}%
   >{\raggedleft\arraybackslash}p{1cm}%
   >{\raggedright\arraybackslash}p{4cm}%
   >{\raggedleft\arraybackslash}p{1cm}%
   >{\raggedright\arraybackslash}p{4cm}%
   >{\raggedleft\arraybackslash}p{1cm}%
   @{}%
  }
\toprule
  \multicolumn{2}{c}{\textbf{Cycling and micromobility}}
& \multicolumn{2}{c}{\textbf{Public transport (short distance)}}
& \multicolumn{2}{c}{\textbf{Public transport (long distance)}} \\
  \textbf{Factor}                  & \textbf{\#} 
& \textbf{Factor}                  & \textbf{\#}
& \textbf{Factor}                  & \textbf{\#} \\
\cmidrule{1-2}\cmidrule{3-4}\cmidrule{5-6}
Road path availability and safety  & $315$ & Privacy                        & $133$ & Today's weather             & $2$ \\
Road path quality                  & $307$ & Other people                   & $111$ & Reliability of travel time  & $2$ \\
Cars other vehicles                & $267$ & Air quality                    & $103$ & Vehicle ride smoothness     & $1$ \\
Air quality                        & $173$ & Noise level                    & $100$ & Ability to do what I wanted & $1$ \\
Noise level                        & $132$ & Crowdedness seating            &   $99$ & Privacy                     & $1$ \\
Road path directness               & $131$ & Seating quality personal space &   $99$ & Other people                & $1$ \\
Today's weather                    & $123$ & Today's weather                &   $84$ & Cleanliness                 & $1$ \\
Traffic signals crossings          & $122$ & Internet connectivity          &   $77$ &                             &     \\
Crowding congestion                & $111$ & Charging opportunity           &   $76$ &                             &     \\
Lighting visibility                &  $88$ & Scenery                        &   $67$ &                             &
\\ \bottomrule
\end{tabular}
\clearpage{}%

\caption{Top negative experience factors for the trip legs selected by our case study.}
\label{tab:selected_experience_factors}
\end{table*}

Table \ref{tab:case-study} presents some statistics related to each step of the process as presented above. For brevity, we will refer to trip legs performed with a mode of transport in the category \emph{private motorized} as legs performed ``by car'' and trips in the categories \emph{cycling and emerging micromobility}, \emph{public transport (short distance)}, \emph{public transport (long distance)} as legs performed ``by alternative modes of transport.''

Tables \ref{tab:all_experience_factors} and \ref{tab:selected_experience_factors} present the results of our case study: Table \ref{tab:all_experience_factors} contains the overall top-ten negative experience factors for all trip legs in the categories \emph{cycling and emerging micromobility}, \emph{public transport (short distance)}, and \emph{public transport (short distance)}. Table \ref{tab:selected_experience_factors} presents the top negative experience factors only for the trip legs select by our case study for the same transport categories.

When cycling we find two main areas of concern: safety (availability of bicycle paths, safety from other cars, visibility and traffic signals) and quality (noise level, air quality). Road path directness has a somewhat a more important role when cycling is used as an alternative to traveling by car w.r.t. the general negative experience factors. Weather is ranked among the top-10 negative experience factors in both bases. For short-distance public transport the main obstacles are lack of privacy and crowdedness in many forms (including noise level and air quality), while reliability of travel time do not appear in the top-10.

\section{Research Opportunities}\label{sec:opportunities}

Our dataset can support numerous research initiatives, with applications that can provide benefits both to the end users and to transport stakeholders, such as transport operators. %

\paragraph{Cost-benefit Analyses} The approach showed in this study challenges conventional cost-benefit narratives and paradigms on the value of travel time, which is the current bedrock of how policy decision in the realm of transport are made. We envision that the ``worthwhile time'' approach can be further developed to become a viable alternative that takes more into account the travelers' perspective, especially when addressing significant infrastructure investments.

\paragraph{User profiling and clustering} Crossing user mobility with their experience factors can be directly used to profile and cluster the users~\cite{BasileCMB20}. The identification of users with similar behavior and a similar value of travel time might directly impact the shaping of journey planners, with solutions that target that cluster being presented first (e.g., if a user belongs to a cluster associated with low emissions, a sorting by emissions might be the default options). %

\paragraph{Recommender systems} Current  systems associate mobility to Point-of-Interest recommendation~\cite{LiuPCY17}. While in this domain collaborative filtering can be enriched with geographical information, our dataset can offer much richer notions of peer users. With the new prospective provided by our analysis, a peer user is not only someone who visited PoI similarly to another, but can be someone who gives values to the same experience factors. In addition, our dataset can enable novel forms of recommendation, based on the previous observations, such as the suggestion of  activities to perform given a type of trip (e.g., reading in trains).

\paragraph{Ad targeting} Most platforms nowadays survive thanks to advertisements, which are usually personalized based on the experience of the users in a platform~\cite{SaiaBCF16}. It is clear that in the context of value of travel time, ad targeting should go beyond this, to consider the factors that positively or negatively impact user mobility, and tailor ads around them. These ads would be much more effective, thus benefiting the platform, and by presenting the users with ``complementary'' services, based on their preferences.

\section{Conclusions and Future Work}\label{sec:conclusions}

Extracting actionable knowledge from user behavior in traveling can help knowing them more and providing them with better services (e.g., personalized rankings). For this reason, it is important not only to monitor the users, but to understand them and the choices they make, according to the value of their travel time. Plus, trips are complex entities, made up of several legs, which might disclose information about user behavior at a finer granularity. 

To accomplish the goal of knowing users, their travel preferences and their value of travel time, we presented a mobility dataset collected via an app and we performed an analysis of factors impacting negatively the usage of cycling and public transport. Our dataset collects information about user mobility, capturing both raw information about trips and their legs (e.g., coordinates, time, and weather), plus information about the worthwhileness of a trip leg. 

To assess the impact that our dataset can have in the real-world, we presented also a use-case to analyze what are experience factors that can negatively impacting the use of public transport, by comparing cycling and walking versus private cars. As our use case has shown, our dataset can provide valuable information to transport operators and service providers, so that their services can be tailored on the needs of their users. Our deliverable also presents opportunities in other domains, such as personalized recommendation or customer clustering. In addition, we plan to use the dataset to take action in concrete real-world scenarios, such as the design of cost-benefit analyses, to see how to improve infrastructures and services, thanks to the needs and preferences of the users. 

\section{Acknowledgements}
This work received the support of the \motiv project, funded from the European Union's Horizon 2020 research and innovation programme under grant agreement No 770145 (\url{https://motivproject.eu/}).
The authors would like to acknowledge the contribution to the project provided by the local data collection managers for their invaluable support in the coordination of the activities in each country.

\bibliography{motiv-dataset}

\end{document}